\documentstyle[epsf,twoside,fleqn,espcrc2]{article}

\title{The $q \bar{q}$ relativistic interaction in the Wilson loop approach}
\author{N. Brambilla$^{\rm a~*}$ and A. Vairo 
\address{Institut f\"ur Theoretische Physik, Universit\"at Heidelberg \\
         Philosophenweg 16, D-69120 Heidelberg, FRG}
\thanks{Alexander von Humboldt Fellow}}

\begin{document}

\begin{abstract}
We study the $q \bar{q}$ relativistic interaction starting from 
the Feynman--Schwinger representation of the gauge-invariant quark-antiquark 
Green function. We focus on the one-body limit and discuss the obtained 
non-perturbative interaction kernel of the Dirac equation.  
\end{abstract}

\maketitle
  
\section{INTRODUCTION}

The dynamics of a system composed by two heavy quarks is  
well understood in terms of a potential interaction (static plus relativistic 
corrections) obtained from the semirelativistic reduction of the QCD dynamics 
\cite{Wilson,BCP,DoSi,DQCD,BV}.  

If at least one of the quarks is light the system behaves relativistically  
(the light quark can no more be considered static) and a pure relativistic 
treatment becomes necessary (via Dirac or Bethe--Salpeter equations).
A lot of phenomenologically justified relativistic equations 
can be found in the literature but up to now we miss a relativistic 
treatment which follows directly from QCD. Our work goes in this direction.  
In order to simplify the problem we focus on the heavy-light mesons 
in the non-recoil limit (i. e. infinitely heavy antiquark). 
Only at the end we will briefly discuss the two-body case. 
Our starting point is the quark-antiquark gauge-invariant 
Green function taken in the infinite mass limit of one particle. 
The only dynamical assumption is on the behaviour of the 
Wilson loop (i. e. on the nature of the non-perturbative vacuum). 
The gauge invariance of the formalism guarantees  
that the relevant physical information are preserved at any step 
of our derivation. In this way we obtain a QCD justified 
fully relativistic interaction kernel for the quark 
in the infinite mass limit of the antiquark. 
This kernel reduces in some region of the physical parameters 
to the heavy quark mass potential, and leads in some other region 
to the heavy quark sum rules results, providing in this 
way an unified description. We discuss our result 
with respect to the old-standing problem of the 
Lorentz structure of the Dirac kernel for a confining interaction
\cite{amerdirac,simdirac,olsdirac}. The main results 
presented here can be found in \cite{bvplb}. 

\section{THE RELATIVISTIC INTERACTION IN THE ONE-BODY LIMIT}

\begin{figure}[htb]
\vskip -1truecm
\makebox[3.5truecm]{\phantom b}
\epsfxsize=7.5truecm
\epsffile{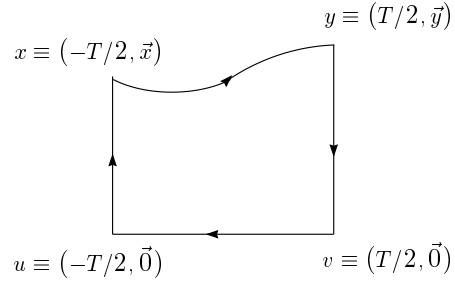}
\vskip -4truecm
\caption{The Wilson loop in the static limit of the heavy quark.}
\label{figwilsonh}
\vskip -0.5truecm
\end{figure}

The quark-antiquark Green function is given in the 
quenched approximation by 
\begin{eqnarray}
&~& G_{\rm inv}(x,u,y,v) =  
\left\langle {\rm Tr} \,i\,S^{(1)}(x,y;A) U(y,v) \right. 
\nonumber\\
&~& \qquad\qquad \times \left. i\,S^{(2)}(v,u;A) U(u,x)\right\rangle, 
\label{Ginv}
\end{eqnarray}
where the points $x,y,u,v$ are defined as in Fig. \ref{figwilsonh},  
$\langle~~\rangle$ means the normalized average over the gauge field  
$A_\mu$,  $S^{(i)}$ is the fermion propagator in the external field $A_\mu$ 
associated with the particle $i$ and the strings 
\begin{eqnarray}
&~& U(y,x) \equiv  
\nonumber\\
&~& \!\!\! {\rm P}\,
\exp \left\{ \displaystyle ig\int_0^1 ds\, (y-x)^\mu  A_\mu(x + s(y-x)) 
\right\}, 
\nonumber
\end{eqnarray}
are needed in order to have gauge invariant initial 
and final bound states. A very convenient way to represent it 
is the so-called Feynman--Schwinger representation 
(see \cite{sitj,fsbs97} and refs. therein), where the fermion propagators 
are expressed in terms of quantomechanical path integrals over the quark 
trajectories ($z_1(t_1)$ and $z_2(t_2)$) 
\begin{eqnarray}
&~&G_{\rm inv}(x,u,y,v) =
{1\over 4} \Bigg\langle {\rm Tr}\,{\rm P}\, 
(i\,{D\!\!\!\!/}_{y}^{\,(1)}+m_1)
\nonumber\\ 
&~& \times \int_{0}^\infty dT_1\int_{x}^{y}{\cal D}z_1
e^{\displaystyle - i\,\int_{0}^{T_1}dt_1 {m^2+\dot z_1^2 \over 2}   }
\nonumber\\
&~& \times
\int_{0}^\infty dT_2\int_{v}^{u}{\cal D}z_2
e^{\displaystyle - i\,\int_{0}^{T_2}dt_2 {m^2+\dot z_2^2 \over 2}   }
\nonumber\\
&~& \times e^{\displaystyle ig \oint_\Gamma dz^\mu A_\mu(z)}
e^{\displaystyle i\,\int_{0}^{T_1}dt_1 {g\over 4}\sigma_{\mu\nu}^{(1)}
F^{\mu\nu}(z_1)}
\nonumber\\
&~& \times
e^{\displaystyle i\,\int_{0}^{T_2}dt_2 {g\over 4}\sigma_{\mu\nu}^{(2)}
F^{\mu\nu}(z_2)} 
\nonumber\\
&~& \times
(-i\,\buildrel{\leftarrow}\over{D\!\!\!\!/}_{v}^{\,(2)} + m_2) \Bigg\rangle . 
\label{Ginv2}
\end{eqnarray} 
From Eq. (\ref{Ginv2}) it emerges quite manifestly that the entire dynamics 
of the system depends on the Wilson loop:
\begin{equation}
W(\Gamma;A) \equiv   
{\rm Tr \,} {\rm P\,} \exp \left\{ ig \oint_\Gamma dz^\mu A_\mu (z) \right\},
\label{wilson}
\end{equation}
being  $\Gamma$ the closed curve defined by the quark trajectories 
and the endpoint strings  $U(y,v)$ and $U(u,x)$.

In order to treat a  simpler case, let us assume, now,  
that the antiquark moving on the second fermion line 
becomes infinitely heavy. The only trajectory surviving 
in the path integral of Eq. (\ref{Ginv2}) associated with the second 
particle is the static straight line propagating from $v$ to $u$. 
The corresponding Wilson loop of the system is represented in 
Fig. \ref{figwilsonh}. As already noted in \cite{bal85} in this case 
it turns out to be convenient to choose the following gauge condition:
\begin{equation}
A_\mu(x_0,{\bf 0}) = 0, \qquad\qquad x^jA_j(x_0,{\bf x}) = 0 .
\label{gauge}
\end{equation} 
Notice that this gauge choice is possible since the formalism 
is completely gauge invariant. Within this gauge it is possible to 
express the gauge field in terms of the field strength tensor, 
$$
A_\mu(x) = \int_0^1 d\alpha\, 
\alpha^{n(\mu)} \,x^k F_{k\mu}(x_0, \alpha {\bf x}), 
$$
where $n(0) = 0$ and $n(i) = 1$.
Moreover the only non-vanishing contribution to the Wilson loop 
is given by the quark paths connecting $x$ with $y$, and we have 
\begin{equation}
W(\Gamma;A) =    
{\rm Tr \,} {\rm P\,} \exp \left\{ ig \int_x^y dz^\mu A_\mu (z) \right\}.
\label{wilson1}
\end{equation}

\begin{figure}[htb]
\vskip -2 truecm
\makebox[7.5truecm]{\phantom b}
\epsfxsize=6truecm
\epsffile{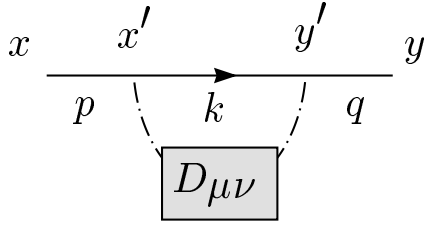}
\vskip -0.5 truecm
\caption{The interaction kernel $K$.}
\label{figfockh}
\vskip -0.5 truecm
\end{figure}

As shown in \cite{sitj,fsbs97} in order to evaluate Eq. (\ref{Ginv2}) 
we need to know the Wilson loop average over the gauge fields. 
We evaluate it via the cumulant expansion described in \cite{DoSi}. 
Keeping only bilocal cumulants we obtain:
\begin{eqnarray}
&~& \langle W(\Gamma,A) \rangle  
\nonumber\\
&~& = 
\exp \left\{ - {g^2\over 2} \int_x^y dx^{\prime\mu} \int_x^y dy^{\prime\nu}
D_{\mu\nu}(x^\prime,y^\prime) \right\},
\nonumber\\
&~& D_{\mu\nu}(x,y) \equiv
x^k y^l\int_0^1 d\alpha \, \alpha^{n(\mu)} 
\nonumber\\
&~&
\times \int_0^1 d\beta\, \beta^{n(\nu)}
\langle F_{k\mu}(x^0,\alpha{\bf x})F_{l\nu}(y^0,\beta{\bf y})\rangle , 
\label{svm}
\end{eqnarray}
Assumption (\ref{svm}) corresponds to the so-called sto\-chastic vacuum 
model. Inserting expression (\ref{svm}) in Eq. (\ref{Ginv2}) 
and expanding the exponential we obtain the following
expression for the propagator $S_D$ of the quark  
(which is $G_{\rm inv}$  ``projected'' on the first fermion line):
\begin{equation}
S_D = S_0 + S_0 \, K \, S_0 + S_0 \, K \, S_0 \, K \, S_0 + \cdots.  
\label{sdexp}
\end{equation} 
$S_0$ is the free fermion propagator. Taking into account only the first 
planar graph (since we are interested only in contributions proportional to 
the gluon condensate), we have $K(y^\prime,x^\prime) = \gamma^\nu 
S_0(y^\prime,x^\prime) \gamma^\mu D_{\mu\nu}(x^\prime,y^\prime)$. 
A graphical representation of $K$ is given in Fig. \ref{figfockh}.
Eq. (\ref{sdexp}) can be written in closed form  
as $S_D = S_0 + S_0 K S_D$ (or in terms of the wave-function, 
$({p\!\!\!/} -m - iK)\psi = 0$; $m\equiv m_1$). Therefore, $K$ can 
be interpreted as the interaction kernel of the Dirac equation associated 
with the motion of a quark in the field generated by an 
infinitely heavy antiquark. 

Assuming that the correlator $\langle F_{\mu\lambda}(x)F_{\nu\rho}(y)\rangle$ 
depends only on the difference between the coordinates, we define: 
\begin{eqnarray}
&~& \langle F_{k\mu}(x^0,\alpha{\bf x})F_{l\nu}(y^0,\beta{\bf y})\rangle 
\nonumber\\
&~&\qquad\quad \equiv f_{k\mu l\nu}(x^0-y^0, \alpha{\bf x} - \beta{\bf y}). 
\nonumber
\end{eqnarray}
With this assumption $K$ can be written in momentum space as: 
\begin{eqnarray}
&~& K(q,p) 
\nonumber\\
&~& = -g^2(2\pi)\delta(p^0-q^0) 
\int_{-\infty}^{+\infty}d\tau \int_0^1 d\alpha \, \alpha^{n(\mu)}
\nonumber\\
&~& \times 
\int_0^1 d\beta \, \beta^{n(\nu)} 
{\partial\over\partial p^k}{\partial\over\partial q^l}
\int d^3r \,e^{i({\bf p}-{\bf q})\cdot{\bf r}} \nonumber\\
&~& \times
\gamma^\nu\left\{ \theta(-\tau)\Lambda_+({\bf t})\gamma^0 e^{-i(p^0-E_t)\tau}
\right.
\nonumber\\
&~& \qquad - 
\left.
\theta(\tau)\Lambda_-({\bf t})\gamma^0 e^{-i(p^0+E_t)\tau} \right\}\gamma^\mu
\nonumber\\
&~& \times
f_{k\mu l\nu}(\tau, (\alpha - \beta){\bf r}),
\label{kmom}
\end{eqnarray}
where ${\bf t} \equiv (\beta{\bf p}-\alpha{\bf q})/(\beta-\alpha)$, 
$E_t = \sqrt{t^2+m^2}$ and $\Lambda_{\pm}({\bf t}) = 
{\displaystyle{E_t \pm (m-{\bf t}\cdot\gamma)\gamma^0 \over 2 E_t}}$.

Equation (\ref{kmom}) is our basic expression. It contains the 
perturbative interaction up to order $g^2$ and the non-perturbative 
one carried by a single insertion of a second order cumulant. 
From now on we want to focus our attention only on the 
purely non-perturbative interaction. The Lorentz structure of the 
non-perturbative relevant part of $f_{\mu\lambda\nu\rho}$ is  
\begin{equation}
f^{\rm n.p.}_{\mu\lambda\nu\rho}(x) = {1 \hskip -2.2 pt {\rm l}}_{\rm c} 
{\langle F^2(0) \rangle \over 24 N_c}
(g_{\mu\nu} g_{\lambda\rho} - g_{\mu\rho} g_{\lambda\nu}) D(x^2)
\label{ddd}
\end{equation}
where  $\langle F^2(0) \rangle$ is the gluon condensate, 
${1 \hskip -2.2 pt {\rm l}}_{\rm c}$ the identity matrix of SU(3) and 
$D$ is a non-per\-turbative form factor normalized to unit at the origin.
Lattice simulations have shown that $D$ falls off exponentially 
(in Euclidean space-time) at long distances with a correlation 
length $a^{-1} \sim (1 ~{\rm GeV})^{-1}$ 
\cite{lat}. As shown in \cite{DoSi} this behaviour of $D$ is sufficient 
to give confinement at least in some kinematic regions. 

We notice here that the most general form of 
$f^{\rm n.p.}_{\mu\lambda\nu\rho}(x)$ could also contain a term of the type 
\begin{eqnarray}
&~& {1 \hskip -2.2 pt {\rm l}}_{\rm c} 
{\langle F^2(0) \rangle \over 24 N_c}
{1\over 2} \bigg\{ \partial_\mu(x_\nu g_{\lambda\rho} - x_\rho 
g_{\nu\lambda}) 
\nonumber\\
&~& \qquad\quad + \partial_\lambda(x_\rho g_{\mu\nu} - x_\nu g_{\mu\rho}) 
\bigg\} D_1(x^2), 
\label{ddd1}
\end{eqnarray}
where $D_1$ is another unknown non-perturbative form factor. 
Since this is the Lorentz structure of the perturbative 
part of $f$, it is not surprising to discover 
that, substituting Eq. (\ref{ddd1}) inside the Wilson loop  
in the large time separation limit, the terms depending on $D_1$ 
give rise to the kernel  
$$
{g^2\over 2} {\langle F^2(0) \rangle \over 24 N_c}
\int_{-\infty}^{+\infty}d\tau \int_0^r d\lambda \lambda 
D_1(\tau^2-\lambda^2)  \,\, \gamma^0,
$$
which is of the Coulomb type. The treatment of this kernel 
is trivial and gives contribution of the perturbative type. 
Therefore in this work we will not consider terms containing $D_1$. 

In what follows we study expression (\ref{kmom})   
for different choices of the parameters which are 
the correlation length $a$, the mass $m$, the binding energy 
$(p^0 - m)$ and the momentum transfer $({\bf p} - {\bf q})$. 
 
{\it A. Heavy quark potential case}  ($m>a>p^0-m$)
\noindent 
If we assume $a$ to be bigger than the binding energy $(p^0 - m)$ 
and smaller than the mass $m$ of the quark, 
since $a\sim 1$ GeV, the quark turns out to be sufficiently heavy 
to be considered non-relativistic. In order to obtain the $1/m^2$ 
potential we can neglect the ``negative  energy states'' 
contributions to (\ref{kmom}) by writing  
\begin{eqnarray}
&~& K(q,p) \simeq -g^2(2\pi)\delta(p^0-q^0) \nonumber\\
&~&  \times \int_0^\infty d\tau \int_0^1 d\alpha \, \alpha^{n(\mu)}
\int_0^1 d\beta \, \beta^{n(\nu)} 
{\partial\over\partial p^k}{\partial\over\partial q^l}
\nonumber\\
&~&  \times \int d^3r e^{i({\bf p}-{\bf q})\cdot{\bf r}} 
\nonumber\\
&~&  \times 
\gamma^\nu \Lambda_+({\bf t})\gamma^0 \gamma^\mu 
f^{\rm n.p.}_{k\mu l\nu}(\tau, (\alpha - \beta){\bf r}). 
\label{kmomA}
\end{eqnarray}
Now, inserting Eq. (\ref{ddd}) and by means of the usual reduction 
techniques, we obtain up to order $1/m^2$ the static and 
spin dependent potential 
\begin{eqnarray}
&~& V(r) 
= g^2 {\langle F^2(0) \rangle \over 24 N_c} 
\int_{-\infty}^{+\infty} d\tau \int_0^r d \lambda  
\nonumber\\
&~&\qquad\qquad 
\times (r-\lambda)\,D(\tau^2 - \lambda^2) 
\nonumber\\
&~& + {{\bf \sigma}\cdot {\bf L} \over 4 m^2} {1\over r}
g^2 {\langle F^2(0) \rangle \over 24 N_c} 
\int_{-\infty}^{+\infty} d\tau \int_0^r d \lambda
\nonumber\\
&~&\qquad\qquad 
\times\left( 2{\lambda\over r} - 1\right) D(\tau^2 - \lambda^2).
\label{pot}
\end{eqnarray}
This result agrees with the one body limit of the potential 
given in \cite{DoSi,BV}. In particular for $r\to\infty$ 
identifying the string tension $\sigma = \displaystyle g^2 
{\langle F^2(0) \rangle \over 24 N_c} \int_{-\infty}^{+\infty} d\tau  
\int_0^\infty d \lambda \, D(\tau^2 - \lambda^2)$ 
we obtain the well-known Eichten and Feinberg result \cite{Wilson}, 
\begin{equation} 
V(r) = \sigma r -C
- {{\bf \sigma}\cdot {\bf L} \over 4 m^2} \,\, {\sigma \over r}, 
\label{EF}
\end{equation}
where $C = \displaystyle g^2 {\langle F^2(0) 
\rangle \over 24 N_c} \int_{-\infty}^{+\infty} d\tau  \int_0^\infty d \lambda 
\, \lambda\, D(\tau^2 - \lambda^2)$. 
We observe that the Lorentz structure  which gives origin to the negative 
sign in front of the spin-orbit potential in (\ref{EF}) is in our 
case not simply a scalar ($K\simeq \sigma \, r$). 

{\it B. Sum rules case}  ($a<p^0-m$, $a<m$)
\noindent
Let us consider now the case in which the binding energy of the quark is 
bigger than the correlation length, which can be considered zero respect 
to all the scales of the problem. We have   
\begin{eqnarray}
&~& K(q,p) \simeq -g^2 (2\pi)\delta(p^0-q^0)
{1 \hskip -2.2 pt {\rm l}}_{\rm c}{ \langle F^2(0) \rangle \over 24 N_c} 
\nonumber\\
&~&\quad
\times(g_{\mu\nu} g_{kl} - g_{\mu l} g_{\nu k}) 
\nonumber\\
&~&\quad 
\times \int_0^1 d\alpha \, \alpha^{n(\mu)}
\int_0^1 d\beta \, \beta^{n(\nu)} 
{\partial\over\partial p^k}{\partial\over\partial q^l}
\nonumber\\
&~&\quad 
\times \left( \gamma^\nu S_0(p) 
\gamma^\mu (2\pi)^3\delta^3({\bf p}-{\bf q})\right). 
\label{kmomB}
\end{eqnarray}
In particular from Eq. (\ref{kmomB}) we obtain the well-known leading 
contribution to the heavy quark condensate \cite{svz}:
\begin{eqnarray}
&~& \langle \bar{Q} Q \rangle  
\nonumber\\
&~& \!\!\!\!\!
= -\int {d^4p \over (2\pi)^4}\int {d^4q \over (2\pi)^4} 
{\rm Tr} \left\{ S_0(q)K(q,p)S_0(p) \right\} 
\nonumber\\
&~&\!\!\!\!\!   
= -{1\over 12} {\langle \alpha F^2(0) \rangle \over \pi m}.
\end{eqnarray}

{\it C. Light quark case}  ($a>m$)
\noindent 
Since we have reproduced the known results concerning heavy quarks, 
Eq. (\ref{kmom}) should maintain some physical meaning also when considering 
heavy-light mesons with a strange quark (like D$_{\rm s}$ and  B$_{\rm s}$). 
In this case the light quark mass is smaller than $a$: $m_{\rm s} \sim$ 
200 MeV $< 1$ GeV. Actually the case $a>m$ 
has to be considered as the only realistic one concerning 
heavy-light mesons. Under this condition either the 
exponent $(p^0 - E_t)$ as well as $(p^0 + E_t)$ can be 
neglected with respect to $a$. Therefore we have:
\begin{eqnarray}
&~& \! K(q,p) \simeq -g^2 (2\pi)\delta(p^0-q^0) 
\nonumber\\
&~& \!
\times \int_0^{+\infty}d\tau \int_0^1 d\alpha \, \alpha^{n(\mu)}
\int_0^1 d\beta \, \beta^{n(\nu)} 
{\partial\over\partial p^k}{\partial\over\partial q^l}
\nonumber\\
&~& \! 
\times \int d^3r e^{i({\bf p}-{\bf q})\cdot{\bf r}} 
\gamma^\nu\left( 
\Lambda_+({\bf t}) - \Lambda_-({\bf t}) \right) \gamma^0 \gamma^\mu
\nonumber\\
&~& \!
\times f^{\rm n.p.}_{k\mu l\nu}(\tau, (\alpha - \beta){\bf r}). 
\label{kmomC}
\end{eqnarray}
We observe that in the zero mass limit this expression 
gives a chirally symmetric interaction (while a purely scalar interaction 
breaks chiral symmetry at any mass scale). This means on one side that our 
interaction keeps the main feature of QCD i. e. in the zero mass limit 
chiral symmetry is broken only spontaneously. On the other side 
this seems to suggest that for very light quarks the projectors $\Lambda_+$ 
and $\Lambda_-$ which appear in (\ref{kmomC}) should be taken 
from the chiral broken solution of the corresponding Dyson--Schwinger 
equation. 

\section{CONCLUSIONS}

In the literature, also recently, a Dirac equation 
with scalar confining kernel (i. e. $K \simeq \sigma \,r$) 
has been used in order to evaluate non-recoil contributions 
to the heavy-light meson spectrum \cite{amerdirac,simdirac,olsdirac}. 
The main argument in favor of this type of kernel is the nature of the 
spin-orbit potential for heavy quarks. This turns out to have a long-range 
vanishing magnetic contribution (according to the Buchm\"uller 
picture of confinement) and is completely described by the Thomas 
precession term. This situation is compatible with a scalar confining  
kernel. However, assuming more sophisticated confinement models with a bigger 
sensitivity to the intermediate distance region, the spin-orbit 
interaction has no more such a simple behaviour. In particular 
  non zero corrections to the magnetic spin-orbit 
potential show up. Moreover, the velocity dependent sector 
of the potential seems not to be compatible with a scalar kernel 
(we refer the reader to \cite{BV} for an exhaustive discussion). 
Therefore also from the point of view of the potential 
theory there are strong indications that the Lorentz structure of the 
confining kernel should be more complicate that a simple scalar one. 
This emerges also in our approach. The kernel (\ref{kmom}) follows 
simply from the assumption on the gauge fields dynamics given 
by Eq. (\ref{svm}) and by taking only one non-perturbative 
gluon insertion on the quark fermion line. 
When performing the potential reduction of this 
kernel in the heavy quark case ({\it A}) we obtain exactly 
the expected  static and spin-dependent  potentials. 
Therefore our conclusion is that there exists at least one 
non scalar  kernel which reproduces for heavy quark 
not only the Eichten and Feinberg potentials in the long distances limit, 
but also the entire stochastic vacuum model spin-dependent potential.  
Moreover when considering $a$, the inverse of the correlation length, small 
with respect to all the energy scales (case {\it B}), the kernel (\ref{kmom}) 
gives back the leading heavy quark sum rules results.  
It is possible to try to extend the range of applicability of 
Eq. (\ref{kmom}) to more realistic cases, like D$_{\rm s}$ and 
B$_{\rm s}$ mesons where the light quark mass is smaller than the 
characteristic correlation length of the two point cumulant (case {\it C}). 
The relevant part of the kernel is also in this case not a simply scalar one. 
We mention that a similar picture of the Lorentz structure of the confining 
Dirac kernel emerges also in the different approach of \cite{swa}. 

An attempt to extend the present approach to the two-body case 
is given in \cite{fsbs97}. The equivalent graphs of Fig. \ref{figfockh}
seem to play a crucial role (in the two-body case 
such kind of graph exists for any fermion line and as exchange graph). 
Nevertheless these graphs are not sufficient in order 
to provide a complete relativistic description of the two-body system. 
The main difficulty is that in this case it does not exist 
a gauge like (\ref{gauge}) which automatically cancels the contributions 
coming from the end-point strings. These contributions 
are necessary in order to restore gauge invariance. 
From this point of view the situation seems to be quite more complicate 
than in the heavy-light system which remains the most natural 
context  to test the formalism.

\end{document}